\documentclass[12pt,a4paper]{article}
\pdfoutput=1
\usepackage[T1]{fontenc}
\usepackage{color}
\usepackage{adjustbox}
\usepackage{amssymb,amsmath,bbold,MnSymbol}
\usepackage{epsf}
\usepackage{epsfig}
\usepackage{relsize}
\usepackage[dvipsnames]{xcolor}
\usepackage[linktoc=page,bookmarks=false,colorlinks=false,linkbordercolor=RoyalBlue,citebordercolor=ForestGreen,urlbordercolor=CornflowerBlue]{hyperref}
\usepackage{latexsym,mathrsfs,dsfont}
\usepackage[normalem]{ulem} 
\usepackage[compress]{cite}
\usepackage{graphicx}
\usepackage{url}
\usepackage{booktabs}
\usepackage{float}
\usepackage{multirow}
\usepackage{changepage}
\usepackage[hypcap]{caption, subcaption}

\usepackage[titles]{tocloft}

\usepackage{tabularx,colortbl}

\usepackage{fancyhdr}
\advance \headheight by 3.0truept       

\allowdisplaybreaks[1]

\definecolor{red}{cmyk}{0,1,1,0.4}
\definecolor{darkgreen}{rgb}{0.0,0.6,0.0}
\definecolor{cDarkGrey}{RGB}{91,91,91}
\definecolor{cGrey}{RGB}{245,243,238}
\definecolor{cBlue}{RGB}{0,110,191}
\definecolor{cLightBlue}{RGB}{214,237,252}
\definecolor{cRed}{RGB}{196,0,100}
\definecolor{cLightRed}{RGB}{254,222,237}
\definecolor{cGreen}{RGB}{0,166,80}
\definecolor{cLightGreen}{RGB}{254,222,237}
\definecolor{cOrange}{RGB}{221,74,44}
\definecolor{cLightOrange}{RGB}{255,215,210}
\definecolor{cPurple}{RGB}{93,35,125}
\definecolor{cLightPurple}{RGB}{241,230,252}
\definecolor{cYellow}{RGB}{252,191,10}
\definecolor{cISSRBlue}{RGB}{0,111,174}
\definecolor{cISSRGrey}{RGB}{167,169,172}

\newcommand{\be}{\begin{equation}}
\newcommand{\ee}{\end{equation}}

\newcounter{TODO}

\def \refsec#1{Section~\ref{#1}}

\newcommand{\DF}{\Delta F}

\newcommand{\GeV}{\,\text{GeV}}
\newcommand{\geV}{\text{GeV}}
\newcommand{\TeV}{\,\text{TeV}}

\newcommand{\KKbar}{K^0-\bar K^0}
\newcommand{\DDbar}{D^0-\bar D^0}
\newcommand{\BBbar}{B_{s,d}-\bar B_{s,d}}

\newcommand{\alS}{\alpha_s}

\newcommand{\muLow}{{\mu_\text{had}}}
\newcommand{\muEW}{{\mu_\text{ew}}}
\newcommand{\muNP}{{\Lambda}}

\newcommand{\wc}[3][{}]{\big[\mathcal{C}_{#2}^{#1}\big]_{#3}}
\newcommand{\wcup}[3][{}]{[{\hat{\mathcal{C}}}_{#2}^{#1}]_{#3}}

\newcommand{\Wc}[2][{}]{\mathcal{C}_{#2}^{#1}}

\newcommand{\Op}[2][{}]{\mathcal{O}_{#2}^{#1}}

\newcommand{\wcL}[3][{}]{[C_{#2}^{#1}]_{#3}}

\newcommand{\OpL}[2][{}]{Q_{#2}^{#1}}
\newcommand{\opL}[3][{}]{[Q_{#2}^{#1}]_{#3}}

\def\as{\alpha_s}
\def\aspi{\frac{\as}{4\pi}}

\begin{document}

\begin{flushleft}
{\em Version of \today}
\end{flushleft}

\vspace{-14mm}
\begin{flushright}
  {AJB-22-5}
\end{flushright}

\medskip

\begin{center}
{\Large\bf\boldmath
  NLO QCD Renormalization Group Evolution for Non-Leptonic $\Delta F=2$  Transitions in the SMEFT
}
\\[1.2cm]
{\bf
  Jason~Aebischer$^{a}$,
  Andrzej~J.~Buras$^{b}$,
  Jacky Kumar$^{b}$
  }\\[0.5cm]

{\small
$^a$Physik-Institut, Universit\"at Z\"urich, CH-8057 Z\"urich, Switzerland \\[0.2cm]
$^b$TUM Institute for Advanced Study,
    Lichtenbergstr. 2a, D-85747 Garching, Germany \\[0.2cm]
}
\end{center}

\vskip 1.0cm

\begin{abstract}
  \noindent
  We present for the first time
  NLO QCD Renormalization Group (RG) evolution matrices for non-leptonic $\Delta F=2$ transitions in the Standard Model Effective Field Theory (SMEFT). To this end we transform first the known two-loop QCD anomalous dimension matrices (ADMs) of the BSM operators in the so-called BMU basis into the ones in the common Weak Effective Theory (WET) basis (the so-called JMS basis) for which
tree-level and one-loop matching to the SMEFT are already known.
This allows us subsequently to find the two-loop QCD ADMs for the SMEFT  non-leptonic $\Delta F=2$  operators in the Warsaw basis. Having all these ingredients we investigate the impact of these NLO QCD effects on the
QCD RG evolution of SMEFT Wilson coefficients for non-leptonic $\Delta F=2$ transitions from the new physics scale $\Lambda$ down to the electroweak scale $\mu_\text{ew}$. The main benefit of these new contributions is that they allow to remove renormalization scheme dependences present both in the one-loop
  matchings between the WET and SMEFT and also between SMEFT and a chosen UV
  completion. But the NLO QCD effects, calculated here in the
  NDR-$\overline{\text{MS}}$ scheme,
  turn out to be small, in the ballpark of  a few percent but larger than one-loop   Yukawa top effects when only the $\Delta F=2$ operators are considered.
The more complicated class of non-leptonic $\Delta F=1$  decays
     will be presented soon in another publication.
\end{abstract}

\setcounter{page}{0}
\thispagestyle{empty}
\newpage

\setcounter{tocdepth}{2}
\setlength{\cftbeforesecskip}{0.21cm}

\newpage

%
%
%
\section{Introduction}
Non-leptonic $\Delta F=2$ transitions, {represented by} $K^0-\bar K^0$, $D^0-\bar D^0$ and
$B^0_{d,s}-\bar B^0_{d,s}$ mixings play very important roles in the tests of the Standard Model (SM) and of
the New Physics (NP) beyond it \cite{Buras:2020xsm}. In order to increase
the precision of these tests it is necessary to go beyond the leading
order (LO) analyses both in the Weak Effective Theory (WET) and also in
the Standard Model Effective Field Theory (SMEFT). To this end it is
mandatory to include first in the renormalization group (RG) analyses in these
theories the one-loop matching contributions, both between these two theories
as well as when passing thresholds at which heavy particles are integrated out.
But this is not the whole story, {a} fact which is forgotten in some
  recent SMEFT analyses present in the literature. To complete a NLO analysis and remove
various renormalization scheme (RS) dependences in the one-loop matching also
two-loop anomalous dimensions of all operators in the WET and SMEFT have
to be included.

The present status of these efforts in the case of non-leptonic {meson} $\Delta F=1$ decays and $\Delta F=2$ quark mixing processes is as follows:
\begin{itemize}
\item
  The matchings in question are known by now both at tree-level \cite{Jenkins:2017jig} and one-loop
  level \cite{Dekens:2019ept}\footnote{Previous partial results can be found, for example, in \cite{Aebischer:2015fzz,Bobeth:2017xry,Hurth:2019ula, Endo:2018gdn,Grzadkowski:2008mf}.}.
\item
  The one-loop ADMs  relevant for the RG in WET \cite{Jenkins:2017dyc,Aebischer:2017gaw} and SMEFT \cite{Jenkins:2013zja,Jenkins:2013wua,Alonso:2013hga} are
  also known.
\item
  The two-loop QCD ADMs relevant for RG evolutions for both    $\Delta F=1$ and $\Delta F=2$ transitions in WET are also known {\cite{Buras:2000if,Aebischer:2021raf,Aebischer:2020dsw}.}
  \end{itemize}

The main goal of the present paper is to  extend the QCD RG evolution in the SMEFT for $\Delta F=2$ transitions beyond the leading order. {In fact at first
  sight this is straightforward because the $SU(3)_c$ symmetry remains unbroken
  in the SMEFT and in the absence of electroweak interactions one could
  just use the NLO QCD BSM analysis of \cite{Buras:2000if} up to the new physics
  scale $\Lambda$ in the so-called BMU operator basis that is useful for NLO QCD calculations. However, in the presence of electroweak interactions, the so-called SMEFT Warsaw basis {\cite{Grzadkowski:2010es}} is more suitable and it is necessary to perform the QCD
  renormalization group analysis within the SMEFT in that basis.
Therefore we present here for the first time two-loop
ADMs for $\Delta F=2$ four-quark operators of the SMEFT. We will demonstrate
that they can be obtained from the two-loop ADMs {in the BMU basis \cite{Buras:2000if}.} As an intermediate step we will present
the ADM relevant for the WET in the so-called JMS basis  \cite{Jenkins:2017jig} which for the matching of the WET to the SMEFT is more useful {than
the BMU basis.}

The main technology presented here can be extended to non-leptonic $\Delta F=1$
transitions but due {to} the large number of operators involved \cite{Buras:2000if,Aebischer:2021raf} the corresponding analysis
is much more complicated and will be presented in due time in a separate publication. Moreover, while in the case of $\Delta F=2$ operators the
  transformation of ADMs from BMU to JMS and SMEFT bases is free from
  contributions  of Fierz-vanishing evanescent operators \cite{Aebischer:2020dsw}, they contribute
  in the $\Delta F=1$ case \cite{Gorbahn:2004my}, which complicates the analysis.

Having all these ingredients we investigate numerically the
  impact of the NLO corrections calculated here
  on the LO RG evolutions of SMEFT Wilson coefficients for $\Delta F=2$ transitions from the new physics scale $\muNP$ down to the electroweak scale $\muEW$
  presented by us in \cite{Aebischer:2020dsw}. Including also
  the effects of top Yukawa couplings at the one-loop level, taken already into account in the latter paper, we find that the main benefit from the
  present analysis is the removal of QCD renormalization scheme dependences
 present currently both in the one-loop
  matchings between the WET and SMEFT and also between SMEFT and a chosen UV
  completion. But the NLO QCD effects in RG QCD evolution, calculated here in the  NDR-$\overline{\text{MS}}$ scheme \cite{Buras:1989xd},
  turn out to be small, in the ballpark of  a few percent but larger than one-loop Yukawa top effects when only the $\Delta F=2$ operators are considered.

Our paper is organized as follows. In \refsec{sec:2}
we derive the general relation between QCD RG evolutions in two different operator bases for the case of $\Delta F=2$ Wilson coefficients. Subsequently
we also give  the corresponding relations for one-loop and two-loop ADMs.
Using these formulae we summarize  in \refsec{sec:3} the one-loop and two-loop
ADMs for $\Delta F=2$ transitions in {the} BMU, JMS and SMEFT bases as well as the matching matrices between
these bases. Subsequently in  \refsec{sec:3a} we analyze numerically the size of the NLO QCD corrections  calculated here by presenting the
  RG evolution matrices in the WET and SMEFT both at LO and NLO in QCD.
  We summarize in
 \refsec{sec:6}

%
%
%
\section{Basic Formulae for NLO QCD RG Evolution}\label{sec:2}
\subsection{Evolution Matrix}
The RG evolution matrix for non-leptonic transitions in the BMU basis for SM and BSM operators is known including one-loop
and two-loop QCD contributions. It is given as follows
\begin{align}
  \label{eq:UBMU}
  \hat U_\text{BMU}(\muLow,\, \muEW) &
  = \left[\hat 1 + \hat J_\text{BMU}\frac{\alS(\muLow)}{4\pi} \right]
    \hat U_\text{BMU}^{(0)}(\muLow,\, \muEW)
    \left[\hat 1 - \hat J_\text{BMU} \frac{\alS(\muEW)}{4\pi} \right]\,,
\end{align}
where $\hat U_\text{BMU}^{(0)}$ is the RS-independent LO evolution
matrix. On the other hand, $ \hat J_\text{BMU}$ stems from the
RS-dependent two-loop ADMs, which makes them sensitive to the
renormalization scheme considered. This scheme dependence is cancelled
by the one of the matching at $\muEW$ and by the one of the hadronic
matrix elements at $\muLow$.
Explicit general expressions for
$\hat U_i^{(0)}$ and $\hat J_i$ in terms of the coefficients of the one-loop and two-loop perturbative expansions for the ADM $\hat\gamma$ and the QCD $\beta$-function can be found including their derivations in Chapter 5 of \cite{Buras:2020xsm}. They will be listed in Section~\ref{EVM}.

It should be stressed that all  two-loop ADMs and the corresponding values of $\hat J_i$ are given in our paper in the NDR-$\overline{\text{MS}}$ scheme as defined in \cite{Buras:1989xd} with
evanescent operators entering two-loop calculations defined by the so-called Greek method. The details in the
context of WET and SMEFT are discussed in Appendix E of \cite{Aebischer:2020dsw}.

Our goal is  to obtain an analogous expression in the SMEFT, that is
\begin{align}
  \label{eq:USMEFT}
  \hat U_\text{SMEFT}(\muEW,\, \muNP) &
  = \left[\hat 1 + \hat J_\text{SMEFT} \frac{\alS(\muEW)}{4\pi} \right]
    \hat U_\text{SMEFT}^{(0)}(\muEW,\, \muNP)
    \left[\hat 1 - \hat J_\text{SMEFT} \frac{\alS(\muNP)}{4\pi} \right].
\end{align}
To this end we notice that this evolution depends only on
$\hat J_\text{SMEFT}$ and  $\hat U_\text{SMEFT}^{(0)}$ without any
explicit dependence on one- and two-loop anomalous dimensions of the SMEFT operators. This gives us a hint that it should be possible to obtain $\hat J_\text{SMEFT}$ and  $\hat U_\text{SMEFT}^{(0)}$  from the known $\hat J_\text{BMU}$ and  $\hat U_\text{BMU}^{(0)}$ without knowing explicitly one-loop and two-loop
ADMs in the SMEFT. It turns out that this is indeed possible but as an intermediate step we should first find $\hat J_\text{JMS}$ and  $\hat U_\text{JMS}^{(0)}$
from $\hat J_\text{BMU}$ and  $\hat U_\text{BMU}^{(0)}$. We will perform this intermediate step in what follows.

In order to reach this goal we first present the general formula which allows
to obtain $\hat J_\text{B}$ and  $\hat U_\text{B}^{(0)}$ in a given basis B from
$\hat J_\text{A}$ and  $\hat U_\text{A}^{(0)}$ in the basis A for which
these two objects, like in the BMU basis, are already known.

Defining the transformation matrix $\hat R$ between these two operator bases through
\be\label{TR}
\vec Q_B=\hat R \vec Q_A\,,\qquad \vec C_A=\hat R^T \vec C_B\,, \qquad \hat R=\hat R_0+\frac{\alS(\mu)}{4\pi}\hat R_1\,,
\ee
one finds then that in the case of $\Delta F=2$ operators, the absence
  of Fierz-vanishing operator contributions in the NDR-$\overline{\text{MS}}$ scheme implies   $\hat R_1={\hat 0}$ \cite{Aebischer:2020dsw}.
A straightforward calculation results then  in {the} relations we were looking for:
\be\label{GeneralAB}
\boxed{\hat J_\text{B}=(\hat R^{-1}_0)^T \hat J_\text{A} \hat R^T_0\,, \qquad
\hat U_\text{B}^{(0)}(\mu_1,\, \mu_2)=(\hat R^{-1}_0)^T\hat U_\text{A}^{(0)}(\mu_1,\mu_2) \hat R^T_0\,,}
\ee

Let us illustrate these formulae on the two cases of interest.

{\bf Example 1}

Here
\be
\text{A=BMU}, \qquad \text{B=JMS}, \qquad \hat R_0^T\equiv\hat M_0
\ee
and we find
\be
\boxed{\hat J_\text{JMS}=\hat M^{-1}_0 \hat J_\text{BMU} \hat M_0\,, \qquad
\hat U_\text{JMS}^{(0)}(\mu_1,\, \mu_2)=\hat M^{-1}_0\hat U_\text{BMU}^{(0)}(\mu_1,\mu_2) \hat M_0\,.}
\ee

{\bf Example 2}

Here
\be
\text{A=JMS}, \qquad \text{B=SMEFT}, \qquad \hat R_0^T\equiv\hat K_0\,.
\ee
and we find
\be\label{SMEFTResults}
\boxed{\hat J_\text{SMEFT}=\hat K^{-1}_0 \hat J_\text{JMS} \hat K_0\,, \qquad
\hat U_\text{SMEFT}^{(0)} =\hat K^{-1}_0\hat U_\text{JMS}^{(0)} \hat K_0\,,}
\ee
where we suppressed the scales in the last relation. The reason is that {the SMEFT
and WET} are valid at different energy scales. Therefore in order to use
the second relation in the equation above in the basic formula (\ref{eq:USMEFT}) one  should properly adjust the scales
in $\hat U_\text{JMS}^{(0)}$. To test it one can use the one-loop ADMs of
  \cite{Jenkins:2013zja,Jenkins:2013wua,Alonso:2013hga}.

  In summary starting with the known RG NLO evolution in the BSM basis one
  can not only find the corresponding NLO evolution in WET in the JMS basis but also the NLO evolution in the SMEFT.

  This formulation is general but in the case at hand{, while
    $\hat R_0\not=\hat 1$, the transformation from the JMS to the SMEFT basis simplifies drastically.}
    With $\hat K_0=\hat 1$ one has
\be\label{SMEFTResults2}
\boxed{\hat J_\text{SMEFT}=\hat J_\text{JMS}\,, \qquad
\hat U_\text{SMEFT}^{(0)} =\hat U_\text{JMS}^{(0)}\,.}
\ee
This  demonstrates the advantage of the JMS basis over the BMU basis
    when the matching to {the} SMEFT is {considered}.

  \subsection{One-Loop and Two-Loop ADMs in SMEFT}\label{sec:3}
Despite the possibility of finding the NLO RG evolution matrix
in  the basis  B from  the one in the basis A without knowing anomalous dimensions in basis B, it is useful to find these anomalous dimensions if
one wants to include subsequently the effects of Yukawa couplings.

{The ADM} has the following perturbative expansion
\begin{equation}\label{gg01}
\hat\gamma(\as)=\aspi\hat\gamma^{(0)} + \left(\aspi\right)^2\hat\gamma^{(1)}+\, \mathcal{O}(\alpha_s^3)\,.
\end{equation}

We find then
\be\label{LONLOgamma}
\hat\gamma^{(0)}_\text{JMS}=\hat R_0\hat\gamma^{(0)}_\text{BMU}\hat R^{-1}_0\,,\qquad
  \hat\gamma^{(1)}_\text{JMS}=\hat R_0\hat\gamma^{(1)}_\text{BMU}\hat R^{-1}_0\,
\ee
and
\be\label{JMS-SMEFT}
\hat \gamma^{(0)}_{\text{SMEFT}}  = \hat \gamma^{(0)}_{\text{JMS}}\,, \,\  \,\ \,\
\hat \gamma^{(1)}_{\text{SMEFT}}  = \hat \gamma^{(1)}_{\text{JMS}}\,.
\ee

Starting then with the ADMs in the BMU basis one can find the ADMs in the JMS basis and subsequently ADMs in the SMEFT. However, there is one subtlety
  to be taken care of{: The} anomalous dimension matrix $\hat\gamma$ used by us
  and in the most literature including \cite{Buras:2020xsm}
  and in particular in the BMU basis in \cite{Buras:2000if} governs
  the RG evolution of the matrix elements of the operators involved, while
  the evolution of the corresponding Wilson coefficients is governed by the transposed
  matrix $\hat\gamma^T$. But the authors that introduced the JMS and SMEFT bases
  decided to denote by $\hat\gamma$ the one which usually would be called
  $\hat\gamma^T$.

  {In what} follows we will within QCD use
    the standard notation so that the RG evolution of operator matrix elements
    will be governed by $\hat\gamma$ and the evolution of Wilson coefficients by $\gamma^T$.
    Therefore  when listing ADMs and $\hat J$ in the JMS and SMEFT bases we
    will use the standard notation used also in the BMU basis in
    \cite{Buras:2000if}. The resulting QCD RG evolution matrices at LO and NLO
    obviously do not depend on this choice. We will also see
      at the end of the next section that the LO
      Yukawa
      effects in the RG evolution can also be incorporated in the presence of NLO QCD
      corrections in a straightforward manner.
    \section{One-Loop and Two-Loop ADMs in BMU, JMS and SMEFT Bases}\label{sec:33}
\subsection{BMU}
We begin with the  BMU basis \cite{Buras:2000if} for which
the complete ADMs at NLO in QCD have been calculated in \cite{Buras:2000if}\footnote{In the so-called SUSY basis this calculations has been performed in \cite{Ciuchini:1997bw}.}
The BMU basis consists in full generality of $(5 + 3) = 8$ physical operators belonging to the five distinct sectors (VLL, SLL, LR, VRR, SRR). However,
SLL and SRR operators, violating hypercharge conservation are not allowed within the SMEFT at dimension-six level and we will not consider them in what follows. {Adopting the \texttt{WCxf} convention \cite{Aebischer:2017ugx},} the remaining four operators are ($ij = sd, db, sb,cu)$\footnote{We use the
  ordering of flavours as in \cite{Aebischer:2020dsw} but different papers use
  different conventions, which has to be taken into account.}
\begin{equation}
  \label{eq:BMU-basis}
\begin{aligned}
  \OpL[ij]{\text{VLL}} &
  = [\bar{d}_i \gamma_\mu P_L d_j][\bar{d}_i \gamma^\mu P_L d_j]\,, & \OpL[ij]{\text{VRR}} &
  = [\bar{d}_i \gamma_\mu P_R d_j][\bar{d}_i \gamma^\mu P_R d_j]\,,
\\[0.2cm]
  \OpL[ij]{\text{LR},1} &
  = [\bar{d}_i \gamma_\mu P_L d_j][\bar{d}_i \gamma^\mu P_R d_j]\,, & \qquad\quad
  \OpL[ij]{\text{LR},2} &
  = [\bar{d}_i P_L d_j][\bar{d}_i P_R d_j]\,,
\end{aligned}
\end{equation}
which are built exclusively out of colour-singlet currents $[\bar{d}^\alpha_i
  \ldots d^\alpha_j] [\bar{d}^\beta_i \ldots d^\beta_j]$,
where $\alpha,\, \beta$ denote colour indices. This feature is very useful for calculations in DQCD \cite{Buras:2018lgu, Aebischer:2018rrz}, because
their matrix elements
in the large-$N_c$ limit can be obtained directly without using Fierz identities.

The one-loop and two-loop anomalous dimensions are given as follows
\be
\hat \gamma^{(0)}_{\text{BMU}} =
\begin{pmatrix}
6 - \frac{6}{N_c} & 0 & 0 & 0 \\[0.2cm]
0 & 6 - \frac{6}{N_c} & 0 & 0 \\[0.2cm]
0 &0 & \frac{6}{N_c} & 12 \\[0.2cm]
0 &0 & 0   & -6N_c + \frac{6}{N_c}
  \end{pmatrix}     \,, \,\
\hat \gamma^{(1)}_{\text{BMU}} =
\begin{pmatrix}
\hat \gamma^{(1)}_{\text{VLL}} &0  & 0 &0 \\[0.2cm]
0 & \hat \gamma^{(1)}_{\text{VRR}} &0 &0 \\[0.2cm]
0 & 0 & {(\hat \gamma^{(1)}_{\text{LR}})_{11}} & {(\hat \gamma^{(1)}_{\text{LR}})_{12}} \\[0.2cm]
0 & 0 & {(\hat \gamma^{(1)}_{\text{LR}})_{21}} & {(\hat \gamma^{(1)}_{\text{LR}})_{22}}
  \end{pmatrix}.
\ee
Here
\be
\hat \gamma^{(1)}_{\text{VLL}} = \hat \gamma^{(1)}_{\text{VRR}} = -\frac{19}{6} N_c -\frac{22}{3}
+\frac{39}{N_c} - \frac{57}{2N_c^2}  + \frac{2}{3} N_f
-\frac{2}{3N_c} N_f\,,
\ee
and
\begin{eqnarray}
{(\hat \gamma^{(1)}_{\text{LR}})_{11}} &= &  \frac{137}{6}+\frac{15}{2N_c^2}-\frac{22}{3N_c} N_f\,,\\
{(\hat \gamma^{(1)}_{\text{LR}})_{12}} &=& \frac{200}{3}N_c -\frac{6}{N_c} -\frac{44}{3}N_f\,, \\
{(\hat \gamma^{(1)}_{\text{LR}})_{21}} & =&  \frac{71}{4} N_c+\frac{9}{N_c} - 2N_f\,, \\
{(\hat \gamma^{(1)}_{\text{LR}})_{22}} &= & -\frac{203}{6}N_c^2 +{\frac{479}{6}}
+ \frac{15}{2N_c^2} + \frac{10}{3} N_c N_f - \frac{22}{3N_c} N_f.
\end{eqnarray}

Here $N_c$ is the number of colours with $N_c=3$ in QCD. $N_f$ is the
  number of quark flavours, $N_f=3,4,5$ in the WET and $N_f=6$ in the SMEFT.
The numerical solutions for evolution matrices for
$ij = sd, db, sb$ are given in \cite{Buras:2001ra}.

\subsection{JMS}

The JMS basis has been introduced to
facilitate the classification of the complete WET operator basis
\cite{Jenkins:2017jig} for the purpose of matching {from SMEFT onto WET}. The relevant $\DF=2$
operators are

\begin{equation}
  \label{eq:JMS-basis}
\begin{aligned}
  \opL[VLL]{dd}{ijij} & = \OpL[ij]{\text{VLL}} \,, \,\  \,\
  \opL[VRR]{dd}{ijij}  = \OpL[ij]{\text{VRR}} \,, \,\  \,\
  \opL[V1,LR]{dd}{ijij}  = \OpL[ij]{\text{LR},1} \,, \\
  \opL[V8,LR]{dd}{ijij} &
  = [\bar{d}_i \gamma_\mu P_L T^A d_j][\bar{d}_i \gamma^\mu P_R T^A d_j]
  =-{\frac{1}{2N_c}} \OpL[ij]{\text{LR},1} -\OpL[ij]{\text{LR},2} \,,
\end{aligned}
\end{equation}
where $T^A$ are $SU(3)_c$ colour generators of the fundamental representation.

Using the relations
above one finds first
\be
\hat R_0 = \hat R_0^{-1}  =\begin{pmatrix}
1 & 0& 0& 0 \\[0.2cm]
0 &1&0& 0 \\[0.2cm]
0 & 0 &1&0 \\[0.2cm]
0 & 0& -\frac{1}{2 N_c}  & -1
 \end{pmatrix}.
\ee

Using subsequently {(\ref{LONLOgamma})  and the results in the BMU basis} we find one-loop and two-loop
ADMs in the JMS basis
\be \label{eq:gammaJMS}
\hat \gamma^{(0)}_{\text{JMS}} =
\begin{pmatrix}
6-\frac{6}{N_c} &0&0&0 \\[0.2cm]
0& 6-\frac{6}{N_c} &0&0\\[0.2cm]
0&0 & 0  &-12  \\[0.2cm]
0&0 & -\frac{6}{N_c} C_F  & -6N_c +\frac{12}{N_c}
\end{pmatrix} \,, \,\
\hat \gamma^{(1)}_{\text{JMS}} =
\begin{pmatrix}
(\hat \gamma_{\text{JMS}}^{(1)})_{11}   &0&0&0 \\[0.2cm]
0& (\hat \gamma_{\text{JMS}}^{(1)})_{22} &0&0\\[0.2cm]
0&0 & (\hat \gamma_{\text{JMS}}^{(1)})_{33}  &(\hat \gamma_{\text{JMS}}^{(1)})_{34}  \\[0.2cm]
0&0 &(\hat \gamma_{\text{JMS}}^{(1)})_{43} &(\hat \gamma_{\text{JMS}}^{(1)})_{44}
\end{pmatrix}\,,
\ee
where we have still used the BMU notation for $\hat\gamma$ so that in
  contrast to \cite{Jenkins:2017dyc}

\begin{equation}
  \mu \frac{d}{d\mu} \vec{C}_{\text{JMS}} =\hat\gamma^T_{\text{JMS}} \,\vec{C}_{\text{JMS}}\,,
\end{equation}
as emphasized above.
Here,
\begin{eqnarray}  \label{eq:gammaJMScom1}
(\hat \gamma_{\text{JMS}}^{(1)})_{11} &=& (\hat \gamma_{\text{JMS}}^{(1)})_{22} =
 {\frac{(N_c-1) \left (171 - 19N_c^2 - N_c(63-4N_f) \right )}{6N_c^2} }  \,, \\
 (\hat \gamma_{\text{JMS}}^{(1)})_{33} &=& \frac{21}{2} (-1 +\frac{1}{N_c^2} ) \,, \\
 (\hat \gamma_{\text{JMS}}^{(1)})_{34} &=& \frac{6}{N_c} -\frac{{200}N_c}{3}+\frac{44N_f}{3}  \,, \\
(\hat \gamma_{\text{JMS}}^{(1)})_{43} &=& \frac{(N_c^2-1)(9-208N_c^2+22 N_c N_f)}{6N_c^3}  \,, \\
(\hat \gamma_{\text{JMS}}^{(1)})_{44} &=& \frac{27+679N_c^2-203N_c^4-88N_cN_f+20N_c^3 N_f}{6N_c^2}  \,, \label{eq:gammaJMScomp3}
\end{eqnarray}
and $C_F= (N_c^2-1)/(2N_c)$.
We have checked that the obtained $\hat \gamma^{(0)}_{\text{JMS}}$ agrees after transposition with the results in \cite{Jenkins:2017dyc}.

\subsection{SMEFT}

\subsubsection{Operators}
In the SMEFT there are five operators which can contribute to $\Delta F=2$
processes at the dimension-six level
\begin{eqnarray}\label{eq:SMEFTops}
\Op[(1)]{qq}  &=& (\bar q_p \gamma_\mu q_r)(\bar q_s \gamma^\mu q_t) \,, \qquad
\Op[(3)]{qq}  = (\bar q_p \gamma_\mu \tau^I q_r)(\bar q_s \gamma^\mu \tau^I q_t) \,, \notag \\
\Op{dd}       &=& (\bar d_p \gamma_\mu d_r)(\bar d_s \gamma^\mu d_t) \,, \qquad
\Op[(1)]{qd}  = (\bar q_p \gamma_\mu q_r)(\bar d_s \gamma^\mu d_t) \,, \notag \\
\Op[(8)]{qd}  &=& (\bar q_p \gamma_\mu T^A q_r)(\bar d_s \gamma^\mu T^A d_t).
\end{eqnarray}

The relevant SMEFT Wilson coefficients  are
\begin{equation}
  \label{eq:wc_smeft_tree}
  B = \left\{
  \Wc[(1)]{qq}{} + \Wc[(3)]{qq}{} \,,\quad
  \Wc[]{aa}{} \,, \quad
 \Wc[(1)]{qa}{} \,,\quad
  \Wc[(8)]{qa}{}   \right\}\, ,
\end{equation}
}
in the down $(a=d)$ and up $(a=u)$ sector, respectively \cite{Aebischer:2015fzz}.
At tree-level for $\BBbar$ and $\KKbar$ mixing one finds the following matching
conditions at~$\muEW$ in the {\bf down-basis}
\hfill
\begin{equation}
  \label{eq:left-smeft-down}
\begin{aligned}
  \wcL[V,LL]{dd}{ijij} &
  = - \wc[(1)]{qq}{ijij} - \wc[(3)]{qq}{ijij} \,, \qquad
&
  \wcL[V,RR]{dd}{ijij}  & = - \wc{dd}{ijij}      \,,
\\
  \wcL[V1,LR]{dd}{ijij} & = - \wc[(1)]{qd}{ijij} \,,
&
  \wcL[V8,RR]{dd}{ijij} & = - \wc[(8)]{qd}{ijij} \,,
\end{aligned}
\end{equation}
and for $\DDbar$ mixing in the {\bf up-basis}
\begin{equation}
  \label{eq:left-smeft-up}
\begin{aligned}
  \wcL[V,LL]{uu}{ijij} &
  = - \wcup[(1)]{qq}{ijij} - \wcup[(3)]{qq}{ijij} \,, \qquad\qquad
&
  \wcL[V,RR]{uu}{ijij}  & = - \wcup{uu}{ijij}      \,,
\\
  \wcL[V1,LR]{uu}{ijij} & = - \wcup[(1)]{qu}{ijij} \,,
&
  \wcL[V8,RR]{uu}{ijij} & = - \wcup[(8)]{qu}{ijij} \,,
\end{aligned}
\end{equation}
{where we have neglected contributions from modified $Z$-coupling operators.}
Note that we use the Hamiltonian for WET to define Wilson coefficients contrary to \cite{Jenkins:2017jig,Dekens:2019ept}, who use the Lagrangian, in consequence minus signs are present in the matching conditions.

{\subsubsection{QCD anomalous dimensions}}
The ADMs in this case are, as given in (\ref{JMS-SMEFT}), the same as for the JMS basis.
Inspecting the RG evolution for $\wcup[(1)]{qq}{ijij}$ and $\wcup[(3)]{qq}{ijij}$ one finds that the sum for $dd$ or $uu$ indices from these two Wilson coefficients evolves without being   affected by other operators and only this sum matches
  on the VLL operator in the JMS basis. While this can be verified
  explicitly by using the RG equations (35) and (36) in \cite{Aebischer:2020dsw},
  the inclusion of flavour diagonal gluon exchanges at {the} two-loop level cannot
  change this property.

\vspace{0.2cm}

\subsubsection{Top Yukawa anomalous dimension}
In this subsection we report the LO ADM resulting from top Yukawa interactions. It reads for a given sector with flavour indices $ij = sd, db, sb,cu$ in the down-basis \cite{Aebischer:2015fzz}:
\begin{equation}\label{YukawaM}
  \hat\gamma_{y_t}^{(0)}=\frac{y_t^2}{16\pi^2}\begin{pmatrix}
  r_{ij} & 0 & 0 & 0 \\[0.2cm]
  0 & 0 & 0 & 0 \\[0.2cm]
  0 &0 &   \frac{1}{2}r_{ij} & 0 \\[0.2cm]
  0 &0 & 0   & \frac{1}{2}r_{ij}
    \end{pmatrix}  \,,
\end{equation}
with the combination $r_{ij}= |V_{ti}|^2+|V_{tj}|^2$. In the up-basis the ADM is zero and the corresponding mixing effects only come into play when considering back-rotation \cite{Aebischer:2020lsx} at the EW scale.

Indeed this anomalous dimension matrix can be extracted from formulae (30)-(33) in \cite{Aebischer:2020dsw}. It is the coefficient of
  \be
  L =  \frac{1}{(4 \pi)^2} \ln \left (\frac{\muEW}{\muNP} \right)\,,
  \ee
  in the solution of RG equations retaining  only the first leading logarithm
  and neglecting the $\mu$-dependence of $y_t$. Defining
  \be
    \wc[(1+3)]{qq}{ijij}\equiv\wc[(1)]{qq}{ijij}+\wc[(3)]{qq}{ijij}\,,
\ee
  these equations
  read in {\bf the down-basis}
  \begin{align}
  \wc[(1+3)]{qq}{ijij}(\muEW) =
  \wc[(1+3)]{qq}{ijij} + y_t^2 & \bigg[
    \lambda_t^{ik} \wc[(1+3)]{qq}{kjij} + \lambda_t^{kj} \wc[(1+3)]{qq}{ikij}\bigg] L \,,
  \label{eq:ll-rge1}
  \\
    \wc[(1)]{qd}{ijij}(\muEW) =
  \wc[(1)]{qd}{ijij} + y_t^2 & \bigg[
      \frac{\lambda_t^{ik}}{2} \wc[(1)]{qd}{kjij}
    + \frac{\lambda_t^{kj}}{2}\wc[(1)]{qd}{ikij}\bigg] L \,,
  \label{eq:ll-rge3}
\\
  \wc[(8)]{qd}{ijij}(\muEW) =
  \wc[(8)]{qd}{ijij} + y_t^2 & \bigg[
     \frac{\lambda_t^{ik}}{2} \wc[(8)]{qd}{kjij}
   + \frac{\lambda_t^{kj}}{2} \wc[(8)]{qd}{ikij}\bigg] L \,,
\label{eq:ll-rge4}
\end{align}
where a summation over $k$ is implied and where we only considered the $\Delta F=2$ operators in \eqref{eq:SMEFTops}. We have suppressed the argument of the NP scale $\muNP$ in the Wilson coefficients and SM parameters on the r.h.s to simplify the notation.
The $\mu$ dependence of $y_t$ will be included in the next subsection.
  \subsection{Explicit Expression for the Evolution Matrix}\label{EVM}
  \subsubsection{Pure QCD}

  We can now find the NLO QCD evolution matrix in any of the bases considered by us
  using the general {formulae} that we recall here in the case of SMEFT for completeness.
Simplifying the notation by dropping the subscript SMEFT we have:
\begin{align}
  \label{eq:USMEFTFINAL}
  \hat U(\muEW,\, \muNP) &
  = \left[1 + \hat J\, \frac{\alS(\muEW)}{4\pi} \right]
    \hat U^{(0)}(\muEW,\, \muNP)
    \left[1 - \hat J\, \frac{\alS(\muNP)}{4\pi} \right].
\end{align}
Here $\hat U^{(0)}(\muEW,\, \muNP)$ denotes the usual  LO RG evolution
matrix that is explicitly given as follows
\begin{equation}\label{u0vd0}
{\hat U^{(0)}(\muEW,\muNP)= \hat V
\left({\left[\frac{\as(\muNP)}{\as(\muEW)}
\right]}^{\frac{\vec\gamma^{(0)}}{2\beta_0}}
   \right)_D \hat V^{-1},}  \end{equation}
\noindent
where $\hat V$ diagonalizes ${\hat\gamma^{(0)T}}$
\begin{equation}\label{ga0d}
{\hat\gamma^{(0)}_D=\hat V^{-1} {\hat\gamma^{(0)T}} \hat V,}
  \end{equation}
\noindent
and $\vec\gamma^{(0)}$ is the vector containing the diagonal elements of
the diagonal matrix {$\hat\gamma^{(0)}_D$}.

The NLO matrix $\hat J$ is given by
\begin{equation}\label{jvs} \hat J=\hat V \hat H \hat V^{-1}
 \end{equation}
with
\begin{equation}\label{sij}
(H)_{ij}=\delta_{ij}(\gamma^{(0)})_i\frac{\beta_1}{2\beta^2_0}-
  \frac{G_{ij}}{2\beta_0+(\gamma^{(0)})_i-
    (\gamma^{(0)})_j}\,, \end{equation}
where
\begin{equation}\label{gvg2}
  \hat G=\hat V^{-1}{\hat\gamma^{(1)T}} \hat V\,,
\end{equation}
with the two-loop matrix $\hat\gamma^{(1)}$ found using (\ref{LONLOgamma})
{and $\beta_1= \frac{34}{3} N_c^2 -\frac{10}{3} N_c N_f - 2C_F N_f$} \cite{Jones:1974mm}.

Setting $N_c=3$ and $N_f=6$ we find


\be
\hat J_{\text {SMEFT}}^{(6)} = \hat J_{\text{JMS}}^{(6)}=
\begin{pmatrix}
1.37 & 0 & 0 & 0\\[0.3cm]
0& 1.37 & 0 & 0 \\[0.3cm]
0&0& 1.47 & 2.75  \\[0.3cm]
0&0& 16.60 & 6.92
\end{pmatrix}\,, \qquad \hat J_{\text {BMU}}^{(6)} =
\begin{pmatrix}
1.37 & 0 & 0 & 0\\[0.3cm]
0& 1.37 & 0 & 0 \\[0.3cm]
0&0&-1.30 &- 1.38  \\[0.3cm]
0&0& -16.60 & 9.69
\end{pmatrix}.
\ee

The corresponding matrices for $N_f=4$ and $N_f=5$ can be found in Appendix~\ref{app:J45}.

\subsubsection{Including Top Yukawa Effects}\label{sec:4}
  Until now we succeeded to find the NLO QCD RG evolution in the SMEFT but
    also the evolution due to the top Yukawa has to be taken into
    account. But
    the two-loop ADM for $\Delta F=2$ operators including top Yukawa couplings is not known at present.   Therefore we can only combine the known LO evolution due to Yukawa couplings   with the NLO QCD evolution just found.
    The full evolution is then given by
\begin{align}
  \label{eq:USMEFTFINALTOT}
  [\hat U(\muEW,\, \muNP)]_{\rm QCD+ y_t} &
  = \left[\hat 1 + \hat J\, \frac{\alS(\muEW)}{4\pi} \right]
    [\hat U^{(0)}(\muEW,\, \muNP)]_{\rm QCD+ y_t}
    \left[\hat 1 - \hat J\, \frac{\alS(\muNP)}{4\pi} \right]\,,
\end{align}
where the label $\rm QCD + y_t$  indicates that besides QCD also Yukawa
contributions have been taken into account. Note that
$\hat J$  contains only
  QCD contributions. As NLO corrections due to top Yukawa
effects are unknown, it is legitimate to proceed in this manner.

There are two routes to find the LO evolution matrix in this formula. If
one is interested only in the numerical result one
 can  simply
   replace the LO QCD evolution matrix in
  (\ref{eq:USMEFTFINAL}) by the one present in the usual computer
   codes {like \texttt{wilson} \cite{Aebischer:2018bkb} or \texttt{DsixTools} \cite{Celis:2017hod,Fuentes-Martin:2020zaz}} for LO RG evolution in the SMEFT.

   However, as demonstrated in \cite{Buras:2018gto} a very accurate analytic
   formula for the RG evolution including scale dependence of $y_t$ can be found. To
   this end we note that
   \be
   [\hat U^{(0)}(\muEW,\, \muNP)]_{\rm QCD+ y_t}=
   [\hat U^{(0)}(\muEW,\, \muNP)]_{\rm QCD}\, [\hat U^{(0)}(\muEW,\, \muNP)]_{\rm  y_t}
\ee
with the QCD evolution matrix given in (\ref{u0vd0}). To find the second matrix
one can simply follow the {\em case C} in Appendix E of
\cite{Buras:2020xsm} that is a particular case of the general formulae in
 \cite{Buras:2018gto}. We find
\be\label{Xm2}
\left[\hat U^{(0)}(\muEW,\, \muNP)\right]_{\rm  y_t}=\text{exp}\left[\frac{1}{8\pi}\frac{\hat{\gamma_t} }{\gamma^{(0)}_m-\beta_0}
\left(\left[\as(\Lambda)\right]^{\frac{\gamma^{(0)}_m}{\beta_0}-1}-
\left[\as(\muEW)\right]^{\frac{\gamma^{(0)}_m}{\beta_0}-1}\right)\right]\,,
\ee
where the diagonal matrix $\hat{\gamma_t}$
\be
\hat{\gamma_t}=\hat b\, y_t^2(\mu_0)\left[\as(\mu_0)\right]^{-\frac{\gamma^{(0)}_m}{\beta_0}}\,,\qquad \gamma^{(0)}_m=8\,,
\ee
with the diagonal matrix $\hat b=\text{diag}(1,0,1/2,1/2)$ deduced from \eqref{YukawaM} and an arbitrary scale $\mu_0$, which we choose to be $160\,\geV$ in our numerical analysis.

\section{Numerical Analysis}\label{sec:3a}
In order to illustrate the importance of NLO QCD corrections within the SMEFT
with respect to the LO ones and top quark Yukawa effects as well as LO and NLO QCD effects within the WET we derive in the following numerical expressions for the evolution matrices in the WET and SMEFT. The results in this section have been obtained using the analytic expressions derived in the previous section.

\subsection{WET}
Setting $\muLow=1.3\GeV$ and $\muEW=160\GeV$ and using the threshold scale $\mu_5=4.2\,\geV$ between $N_f=5$ and $N_f=4$ we find

\be
[\hat U^{(0)}_{\rm JMS} ]_{\rm QCD}  =
\begin{pmatrix}
0.76 & 0 & 0 & 0\\[0.3cm]
0& 0.76 & 0 & 0 \\[0.3cm]
0&0& 1.10 & 0.31  \\[0.3cm]
0&0& 1.38 & 2.71
\end{pmatrix}\,, \quad  [{\hat U}_{\rm JMS} ]_{\rm QCD} =
\begin{pmatrix}
0.76 & 0 & 0 & 0\\[0.3cm]
0&  0.76 & 0 & 0 \\[0.3cm]
0&0& 1.24 & 0.57  \\[0.3cm]
0&0& 2.02 &  3.59
\end{pmatrix}.
\ee

Here $\hat U^{(0)}_{\rm JMS}$ and $\hat U_{\rm JMS}$ are LO and NLO evolution matrices in the JMS basis, respectively. We observe that in the LR sector the NLO
effects are large and it is mandatory to include them in any phenomenological analysis.
\subsection{SMEFT}
Here we study the evolution between $\muEW=160\GeV$ and $\muNP=10\TeV$
for various cases in the $N_f=6$ flavour theory.
\subsubsection{Pure QCD Evolution}

\begin{equation}
[\hat U^{(0)}_{\rm SMEFT} ]_{\rm QCD}  = \begin{pmatrix}
0.89 & 0 & 0 & 0\\[0.3cm]
0& 0.89 & 0 & 0 \\[0.3cm]
0&0& 1.02 & 0.10  \\[0.3cm]
0&0& 0.43 & 1.52
\end{pmatrix}\,, \quad
[\hat U_{\rm SMEFT} ]_{\rm QCD} = \begin{pmatrix}
0.89 & 0  & 0 & 0 \\[0.3cm]
0 &  0.89  & 0 & 0 \\[0.3cm]
0&0& 1.02 & 0.12 \\[0.3cm]
0&0& 0.46 &  1.57
\end{pmatrix}.
\end{equation}

As expected, due to a much slower running of $\alpha_s$ and its smaller value
than in WET, QCD effects are significantly smaller and this applies in particular to NLO QCD effects.

\subsubsection{Pure Yukawa Evolution}
{When only the Yukawa running in the SMEFT at one-loop is considered, the resulting evolution matrix reads}

\begin{equation}
[\hat U^{(0)}_{\rm SMEFT} ]_{\rm y_t} = \begin{pmatrix}
0.98 & 0 & 0 & 0 \\[0.3cm]
0 & 1.00 & 0 & 0 \\[0.3cm]
0&0& 0.99 & 0 \\[0.3cm]
0 & 0 & 0 & 0.99
\end{pmatrix}
\end{equation}
where we have used $y_t(\mu=160 \geV)=0.94$.

\subsubsection{QCD+Yukawa Evolution}
{In this subsection we consider the combination of QCD and Yukawa running effects at LO and NLO. The corresponding evolution matrices read}

\begin{equation}
[\hat U^{(0)}_{\rm SMEFT} ]_{\rm QCD+ y_t}  = \begin{pmatrix}
0.87 & 0 & 0 & 0 \\[0.3cm]
 0 & 0.89 & 0 & 0 \\[0.3cm]
 0 & 0 & 1.01 & 0.10 \\[0.3cm]
 0 & 0 & 0.43 & 1.51
\end{pmatrix}\,, \quad
[\hat U_{\rm SMEFT} ]_{\rm QCD + y_t} = \begin{pmatrix}
0.88 & 0 & 0 & 0 \\[0.3cm]
0 & 0.89 & 0 & 0 \\[0.3cm]
0 & 0 & 1.01 & 0.11 \\[0.3cm]
0 & 0 & 0.44 & 1.54 \\
\end{pmatrix} \,,
\end{equation}

\noindent
where for both matrices we have used eq.~\eqref{eq:USMEFTFINALTOT}, setting $\hat J$ to zero in the LO case, and keeping $\hat J$ in the computation of $\hat U_{\rm SMEFT}$.

We observe that when only $\Delta F=2$ operators are considered the
  impact of top-Yukawa effects is very small. It is significantly larger
  when also $\Delta F=1$ operators are included in the analysis.
  A detailed discussion of these effects has been presented in \cite{Aebischer:2020dsw}.

\section{Conclusions}\label{sec:6}

 The main results of our paper are as follows.
\begin{itemize}
\item
  General formulae for the relation of QCD RG evolution matrices at LO and NLO
  for $\Delta F=2$ Wilson coefficients between two different operator bases{, given in eq.~}\eqref{GeneralAB}.
\item
  The relation of QCD one-loop and two-loop anomalous dimension matrices   between BMU and JMS bases reported in \eqref{LONLOgamma} and JMS and SMEFT bases
  in eq.~\eqref{JMS-SMEFT}.
\item
  The two-loop QCD ADMs  for $\Delta F=2$  operators in the
  JMS WET basis and in the SMEFT Warsaw basis. They are given in
  {eqs.~}\eqref{eq:gammaJMS}-\eqref{eq:gammaJMScomp3} and eq.~\eqref{JMS-SMEFT}.
  \item
 Master formulae for QCD RG evolution matrices for the Wilson coefficients of
 $\Delta F=2$ operators in the SMEFT Warsaw basis at the NLO were derived. They are given in eq.~\eqref{eq:USMEFTFINAL}.
\item
  Generalization of these formulae to include top Yukawa effects at the one-loop
  level.
\end{itemize}
{These findings allow for a general and scheme-independent QCD analysis of non-leptonic $\Delta F=2$ processes in the SMEFT and WET at NLO.}
In a given UV completion in which the Wilson coefficients have been calculated
at a NP scale $\muNP$, our master formulae allow to calculate them
at the $\muEW$ scale.
The inclusion of NLO  QCD corrections in the RG evolution
in the WET from the hadronic scale $\muLow$ to the
electroweak scale $\muEW$  allows a correct matching of Wilson coefficients to the matrix elements
calculated by lattice QCD (LQCD) or other non-perturbative methods sensitive
to renormalization scheme dependences. The use of the JMS basis on the other
hand allows to generalize this formula to the SMEFT, because in this basis the tree-level matching of
SMEFT {onto} WET \cite{Jenkins:2017jig} and the one-loop matching
\cite{Dekens:2019ept, Aebischer:2015fzz} are known.

The main messages from the numerical analysis in Section~\ref{sec:3a} are
as follows:
\begin{itemize}
\item
  The NLO QCD corrections to the $\Delta F=2$ RG evolution matrices within WET
  are substantial.
\item The NLO QCD corrections to the $\Delta F=2$ RG evolution matrices within SMEFT are small, in the ballpark of a few percent.
\item
  Even smaller are top-Yukawa effects if only $\Delta F=2$ operators are included in the analysis.
  \end{itemize}

The small NLO QCD corrections to the $\Delta F=2$ matrices within the SMEFT
could be considered at first sight disappointing. But it should be realized
that they have been calculated in the NDR scheme but could be larger in a different
RS. This also does not preclude significant one-loop
matching contributions, which in principle could be larger than the QCD effects found here.
However, to combine these one-loop matching conditions with the NLO effects calculated
here they have to be calculated in the NDR scheme. Only then RS independent results can be obtained.

\section*{Acknowledgements}

We thank Christoph Bobeth for useful discussions and comments on the document. J.\,A.\ acknowledges financial support from the European Research Council (ERC) under the European Union’s Horizon 2020 research and innovation programme under grant agreement 833280 (FLAY), and from the Swiss National Science Foundation (SNF) under contract 200020-204428.
A.J.B acknowledges financial support from the Excellence Cluster ORIGINS,
funded by the Deutsche Forschungsgemeinschaft (DFG, German Research Foundation)
under Germany´s Excellence Strategy – EXC-2094 – 390783311.
J.K. is financially supported by the Alexander von Humboldt Foundation's
postdoctoral research fellowship.

%
%
%

\appendix

{\boldmath
\section{$\hat J_{\rm JMS}^{(N_f)} $ and $ \hat J_{\rm BMU}^{(N_f)} $}
\label{app:J45}}
For convenience we report in this appendix the $\hat J$ matrices for $N_f=5,4$ flavours in the JMS and BMU basis, obtained from eq.~\eqref{jvs}.

\begin{align}
\hat J_{\text{JMS}}^{(5)} &=
\begin{pmatrix}
1.63 & 0 & 0 & 0\\[0.3cm]
0& 1.63 & 0 & 0 \\[0.3cm]
0&0& 1.67 & 2.44  \\[0.3cm]
0&0& 17.04 & 5.12
\end{pmatrix}\,, &\qquad \hat J_{\text {JMS}}^{(4)} &= \begin{pmatrix}
1.79 & 0 & 0 & 0\\[0.3cm]
0& 1.79 & 0 & 0 \\[0.3cm]
0&0& 2.43 & 2.10  \\[0.3cm]
0&0& 21.15 & 3.18
\end{pmatrix}\,,\\
\hat J_{\text{BMU}}^{(5)} &=
\begin{pmatrix}
1.63 & 0 & 0 & 0\\[0.3cm]
0& 1.63 & 0 & 0 \\[0.3cm]
0&0& -1.17 & -1.39  \\[0.3cm]
0&0& -17.04 & 7.96
\end{pmatrix}\,,& \qquad \hat J_{\text {BMU}}^{(4)} &= \begin{pmatrix}
1.79 & 0 & 0 & 0\\[0.3cm]
0& 1.79 & 0 & 0 \\[0.3cm]
0&0& -1.10 & -1.39  \\[0.3cm]
0&0& -21.15 & 6.71
\end{pmatrix}\,.
\end{align}

%
%

\renewcommand{\refname}{R\lowercase{eferences}}

\addcontentsline{toc}{section}{References}

\bibliographystyle{JHEP}

\small

\bibliography{Bookallrefs}

\providecommand{\href}[2]{#2}\begingroup\raggedright\begin{thebibliography}{10}

\bibitem{Buras:2020xsm}
A.~J. Buras, {\em {Gauge Theory of Weak Decays}}.
\newblock Cambridge University Press, 6, 2020.

\bibitem{Jenkins:2017jig}
E.~E. Jenkins, A.~V. Manohar, and P.~Stoffer, {\it {Low-Energy Effective Field
  Theory below the Electroweak Scale: Operators and Matching}},  {\em JHEP}
  {\bf 03} (2018) 016, [\href{http://arxiv.org/abs/1709.04486}{{\tt
  arXiv:1709.04486}}].

\bibitem{Dekens:2019ept}
W.~Dekens and P.~Stoffer, {\it {Low-energy effective field theory below the
  electroweak scale: matching at one loop}},  {\em JHEP} {\bf 10} (2019) 197,
  [\href{http://arxiv.org/abs/1908.05295}{{\tt arXiv:1908.05295}}].

\bibitem{Aebischer:2015fzz}
J.~Aebischer, A.~Crivellin, M.~Fael, and C.~Greub, {\it {Matching of gauge
  invariant dimension-six operators for $b\to s$ and $b\to c$ transitions}},
  {\em JHEP} {\bf 05} (2016) 037, [\href{http://arxiv.org/abs/1512.02830}{{\tt
  arXiv:1512.02830}}].

\bibitem{Bobeth:2017xry}
C.~Bobeth, A.~J. Buras, A.~Celis, and M.~Jung, {\it {Yukawa enhancement of
  $Z$-mediated new physics in $\Delta S = 2$ and $\Delta B = 2$ processes}},
  {\em JHEP} {\bf 07} (2017) 124, [\href{http://arxiv.org/abs/1703.04753}{{\tt
  arXiv:1703.04753}}].

\bibitem{Hurth:2019ula}
T.~Hurth, S.~Renner, and W.~Shepherd, {\it {Matching for FCNC effects in the
  flavour-symmetric SMEFT}},  {\em JHEP} {\bf 06} (2019) 029,
  [\href{http://arxiv.org/abs/1903.00500}{{\tt arXiv:1903.00500}}].

\bibitem{Endo:2018gdn}
M.~Endo, T.~Kitahara, and D.~Ueda, {\it {SMEFT top-quark effects on $\Delta
  F=2$ observables}},  {\em JHEP} {\bf 07} (2019) 182,
  [\href{http://arxiv.org/abs/1811.04961}{{\tt arXiv:1811.04961}}].

\bibitem{Grzadkowski:2008mf}
B.~Grzadkowski and M.~Misiak, {\it {Anomalous Wtb coupling effects in the weak
  radiative B-meson decay}},  {\em Phys.~Rev.} {\bf D78} (2008) 077501,
  [\href{http://arxiv.org/abs/0802.1413}{{\tt arXiv:0802.1413}}].

\bibitem{Jenkins:2017dyc}
E.~E. Jenkins, A.~V. Manohar, and P.~Stoffer, {\it {Low-Energy Effective Field
  Theory below the Electroweak Scale: Anomalous Dimensions}},  {\em JHEP} {\bf
  01} (2018) 084, [\href{http://arxiv.org/abs/1711.05270}{{\tt
  arXiv:1711.05270}}].

\bibitem{Aebischer:2017gaw}
J.~Aebischer, M.~Fael, C.~Greub, and J.~Virto, {\it {B physics Beyond the
  Standard Model at One Loop: Complete Renormalization Group Evolution below
  the Electroweak Scale}},  {\em JHEP} {\bf 09} (2017) 158,
  [\href{http://arxiv.org/abs/1704.06639}{{\tt arXiv:1704.06639}}].

\bibitem{Jenkins:2013zja}
E.~E. Jenkins, A.~V. Manohar, and M.~Trott, {\it {Renormalization Group
  Evolution of the Standard Model Dimension Six Operators I: Formalism and
  lambda Dependence}},  {\em JHEP} {\bf 10} (2013) 087,
  [\href{http://arxiv.org/abs/1308.2627}{{\tt arXiv:1308.2627}}].

\bibitem{Jenkins:2013wua}
E.~E. Jenkins, A.~V. Manohar, and M.~Trott, {\it {Renormalization Group
  Evolution of the Standard Model Dimension Six Operators II: Yukawa
  Dependence}},  {\em JHEP} {\bf 01} (2014) 035,
  [\href{http://arxiv.org/abs/1310.4838}{{\tt arXiv:1310.4838}}].

\bibitem{Alonso:2013hga}
R.~Alonso, E.~E. Jenkins, A.~V. Manohar, and M.~Trott, {\it {Renormalization
  Group Evolution of the Standard Model Dimension Six Operators III: Gauge
  Coupling Dependence and Phenomenology}},  {\em JHEP} {\bf 04} (2014) 159,
  [\href{http://arxiv.org/abs/1312.2014}{{\tt arXiv:1312.2014}}].

\bibitem{Buras:2000if}
A.~J. Buras, M.~Misiak, and J.~Urban, {\it {Two loop QCD anomalous dimensions
  of flavor changing four quark operators within and beyond the standard
  model}},  {\em Nucl.~Phys.} {\bf B586} (2000) 397--426,
  [\href{http://arxiv.org/abs/hep-ph/0005183}{{\tt hep-ph/0005183}}].

\bibitem{Aebischer:2021raf}
J.~Aebischer, C.~Bobeth, A.~J. Buras, J.~Kumar, and M.~Misiak, {\it {General
  non-leptonic \ensuremath{\Delta}F = 1 WET at the NLO in QCD}},  {\em JHEP}
  {\bf 11} (2021) 227, [\href{http://arxiv.org/abs/2107.10262}{{\tt
  arXiv:2107.10262}}].

\bibitem{Aebischer:2020dsw}
J.~Aebischer, C.~Bobeth, A.~J. Buras, and J.~Kumar, {\it {SMEFT ATLAS of
  $\Delta$F = 2 transitions}},  {\em JHEP} {\bf 12} (2020) 187,
  [\href{http://arxiv.org/abs/2009.07276}{{\tt arXiv:2009.07276}}].

\bibitem{Grzadkowski:2010es}
B.~Grzadkowski, M.~Iskrzynski, M.~Misiak, and J.~Rosiek, {\it {Dimension-Six
  Terms in the Standard Model Lagrangian}},  {\em JHEP} {\bf 1010} (2010) 085,
  [\href{http://arxiv.org/abs/1008.4884}{{\tt arXiv:1008.4884}}].

\bibitem{Gorbahn:2004my}
M.~Gorbahn and U.~Haisch, {\it {Effective Hamiltonian for non-leptonic $|\Delta
  F| = 1$ decays at NNLO in QCD}},  {\em Nucl.~Phys.} {\bf B713} (2005)
  291--332, [\href{http://arxiv.org/abs/hep-ph/0411071}{{\tt hep-ph/0411071}}].

\bibitem{Buras:1989xd}
A.~J. Buras and P.~H. Weisz, {\it {QCD Nonleading Corrections to Weak Decays in
  Dimensional Regularization and 't Hooft-Veltman Schemes}},  {\em Nucl. Phys.}
  {\bf B333} (1990) 66--99.

\bibitem{Ciuchini:1997bw}
M.~Ciuchini, E.~Franco, V.~Lubicz, G.~Martinelli, I.~Scimemi, et~al., {\it
  {Next-to-leading order QCD corrections to $\Delta F = 2$ effective
  Hamiltonians}},  {\em Nucl.~Phys.} {\bf B523} (1998) 501--525,
  [\href{http://arxiv.org/abs/hep-ph/9711402}{{\tt hep-ph/9711402}}].

\bibitem{Aebischer:2017ugx}
J.~Aebischer et~al., {\it {WCxf: an exchange format for Wilson coefficients
  beyond the Standard Model}},  {\em Comput. Phys. Commun.} {\bf 232} (2018)
  71--83, [\href{http://arxiv.org/abs/1712.05298}{{\tt arXiv:1712.05298}}].

\bibitem{Buras:2018lgu}
A.~J. Buras and J.-M. G{\'e}rard, {\it {Dual QCD Insight into BSM Hadronic
  Matrix Elements for $K^0-\bar K^0$ Mixing from Lattice QCD}},
  \href{http://arxiv.org/abs/1804.02401}{{\tt arXiv:1804.02401}}.

\bibitem{Aebischer:2018rrz}
J.~Aebischer, A.~J. Buras, and J.-M. G{\'e}rard, {\it {BSM hadronic matrix
  elements for $\epsilon'/\epsilon$ and $K\to\pi\pi$ decays in the Dual QCD
  approach}},  {\em JHEP} {\bf 02} (2019) 021,
  [\href{http://arxiv.org/abs/1807.01709}{{\tt arXiv:1807.01709}}].

\bibitem{Buras:2001ra}
A.~J. Buras, S.~J{\"a}ger, and J.~Urban, {\it {Master formulae for $\Delta F=2$
  NLO QCD factors in the standard model and beyond}},  {\em Nucl.~Phys.} {\bf
  B605} (2001) 600--624, [\href{http://arxiv.org/abs/hep-ph/0102316}{{\tt
  hep-ph/0102316}}].

\bibitem{Aebischer:2020lsx}
J.~Aebischer and J.~Kumar, {\it {Flavour Violating Effects of Yukawa Running in
  SMEFT}},  {\em JHEP} {\bf 09} (2020) 187,
  [\href{http://arxiv.org/abs/2005.12283}{{\tt arXiv:2005.12283}}].

\bibitem{Jones:1974mm}
D.~R.~T. Jones, {\it {Two Loop Diagrams in Yang-Mills Theory}},  {\em Nucl.
  Phys. B} {\bf 75} (1974) 531.

\bibitem{Aebischer:2018bkb}
J.~Aebischer, J.~Kumar, and D.~M. Straub, {\it {Wilson: a Python package for
  the running and matching of Wilson coefficients above and below the
  electroweak scale}},  {\em Eur. Phys. J.} {\bf C78} (2018), no.~12 1026,
  [\href{http://arxiv.org/abs/1804.05033}{{\tt arXiv:1804.05033}}].

\bibitem{Celis:2017hod}
A.~Celis, J.~Fuentes-Martin, A.~Vicente, and J.~Virto, {\it {DsixTools: The
  Standard Model Effective Field Theory Toolkit}},  {\em Eur. Phys. J.} {\bf
  C77} (2017), no.~6 405, [\href{http://arxiv.org/abs/1704.04504}{{\tt
  arXiv:1704.04504}}].

\bibitem{Fuentes-Martin:2020zaz}
J.~Fuentes-Martin, P.~Ruiz-Femenia, A.~Vicente, and J.~Virto, {\it {DsixTools
  2.0: The Effective Field Theory Toolkit}},  {\em Eur. Phys. J. C} {\bf 81}
  (2021), no.~2 167, [\href{http://arxiv.org/abs/2010.16341}{{\tt
  arXiv:2010.16341}}].

\bibitem{Buras:2018gto}
A.~J. Buras and M.~Jung, {\it {Analytic inclusion of the scale dependence of
  the anomalous dimension matrix in Standard Model Effective Theory}},  {\em
  JHEP} {\bf 06} (2018) 067, [\href{http://arxiv.org/abs/1804.05852}{{\tt
  arXiv:1804.05852}}].

\end{thebibliography}\endgroup

\end{document}